\documentstyle[12pt]{article}
\def\Re{{\cal R \mskip-4mu \lower.1ex \hbox{\it e}\,}}
\def\Im{{\cal I \mskip-5mu \lower.1ex \hbox{\it m}\,}}
\def\ie{{\it i.e.}}
\def\eg{{\it e.g.}}

\def\etal{{\it et al.}}

\def\sub#1{_{\lower.25ex\hbox{$\scriptstyle#1$}}}
\def\sul#1{_{\kern-.1em#1}}
\def\sll#1{_{\kern-.2em#1}}
\def\sbl#1{_{\kern-.1em\lower.25ex\hbox{$\scriptstyle#1$}}}
\def\ssb#1{_{\lower.25ex\hbox{$\scriptscriptstyle#1$}}}
\def\sbb#1{_{\lower.4ex\hbox{$\scriptstyle#1$}}}

\def\to{\rightarrow}
\def\mh{\ifmmode m\sbl H \else $m\sbl H$\fi}
\def\mch{\ifmmode m_{H^\pm} \else $m_{H^\pm}$\fi}
\def\mt{\ifmmode m_t\else $m_t$\fi}
\def\mc{\ifmmode m_c\else $m_c$\fi}
\def\mz{\ifmmode M_Z\else $M_Z$\fi}
\def\mw{\ifmmode M_W\else $M_W$\fi}
\def\mws{\ifmmode M_W^2 \else $M_W^2$\fi}
\def\mhs{\ifmmode m_H^2 \else $m_H^2$\fi}
\def\mzs{\ifmmode M_Z^2 \else $M_Z^2$\fi}
\def\mts{\ifmmode m_t^2 \else $m_t^2$\fi}
\def\mcs{\ifmmode m_c^2 \else $m_c^2$\fi}
\def\mchs{\ifmmode m_{H^\pm}^2 \else $m_{H^\pm}^2$\fi}
\def\ztwo{\ifmmode Z_2\else $Z_2$\fi}
\def\zone{\ifmmode Z_1\else $Z_1$\fi}
\def\mtwo{\ifmmode M_2\else $M_2$\fi}
\def\mone{\ifmmode M_1\else $M_1$\fi}
\def\tb{\ifmmode \tan\beta \else $\tan\beta$\fi}
\def\xw{\ifmmode x\sub w\else $x\sub w$\fi}
\def\ch{\ifmmode H^\pm \else $H^\pm$\fi}
\def\lum{\ifmmode {\cal L}\else ${\cal L}$\fi}
\def\inpb{\ifmmode {\rm pb}^{-1}\else ${\rm pb}^{-1}$\fi}
\def\infb{\ifmmode {\rm fb}^{-1}\else ${\rm fb}^{-1}$\fi}
\def\epem{\ifmmode e^+e^-\else $e^+e^-$\fi}
\def\ppb{\ifmmode \bar pp\else $\bar pp$\fi}

\def\bsg{\ifmmode b\rightarrow s\gamma \else $b\rightarrow s\gamma$\fi}

\newskip\zatskip \zatskip=0pt plus0pt minus0pt
\def\matth{\mathsurround=0pt}

\def\atversim#1#2{\lower0.7ex\vbox{\baselineskip\zatskip\lineskip\zatskip
  \lineskiplimit 0pt\ialign{$\matth#1\hfil##\hfil$\crcr#2\crcr\sim\crcr}}}

\renewcommand{\thefootnote}{\fnsymbol{footnote}}

\begin{document} \begin{titlepage}
\setcounter{page}{1}
\thispagestyle{empty}
\rightline{\vbox{\halign{&#\hfil\cr
&SLAC-PUB-6579\cr
&July 1994\cr
&T/E\cr}}}
\vspace{0.8in}
\begin{center}

{\Large\bf
Diphoton Production at Hadron Colliders and New Contact Interactions}
\footnote{Work supported by the Department of
Energy, contract DE-AC03-76SF00515.}
\medskip

\normalsize THOMAS G. RIZZO
\\ \smallskip
{\it {Stanford Linear Accelerator Center\\Stanford University,
Stanford, CA 94309}}\\

\end{center}

\begin{abstract}

We explore the capability of the Tevatron and LHC to place limits on the
possible existence of flavor-independent $q \bar q \gamma\gamma$ contact
interactions which can lead to an excess of diphoton events with large
invariant masses. Assuming no departure from the Standard Model is
observed, we show that the Tevatron will eventually be able to place a lower
bound of 0.5-0.6 TeV on the scale associated with this new contact
interaction. At the LHC, scales as large as 3-6 TeV may be probed with
suitable detector cuts and an integrated luminosity of $100 fb^{-1}$.

\end{abstract}

\vskip0.45in
\begin{center}

Submitted to Physical Review {\bf D}.

\end{center}

%\noindent{(Talk given at the {\it Workshop on Photon Radiation from Quarks},
%Annecy, France, December 2-3, 1991.)}

\renewcommand{\thefootnote}{\arabic{footnote}} \end{titlepage}

%%%%%%%%%%%%%%%%%%%%%%%%%%%%%%%---- text

Although the Standard Model(SM) appears to be as healthy as ever{\cite {mor}},
it is generally believed that new physics must exist to address all of the
questions the SM leaves unanswered and which can explain the values of the
various input parameters (\eg, fermion masses and mixing angles). Although
there are many suggestions in the literature, no one truly knows the form
this new physics might take or how it may first manifest itself. Instead of
the direct production of new particles, physics beyond the SM may first
appear as deviations in observables away from SM expectations, such as in the
rates for rare processes or in precision electroweak tests. Another
possibility is that deviations in cross sections of order unity may
be observed once sufficiently high energy scales are probed. This kind of new
physics can generally be parameterized via a finite set of non-renormalizable
contact interactions, an approach which is quite popular in the
literature{\cite {contact}}. In fact, limits already exist from a number of
experiments on the scales associated with contact interactions of various
types{\cite {lims}}.

In this paper we will explore the capability of both the Tevatron and the CERN
Large Hadron Collider(LHC) to probe the existence of flavor-independent(apart
from electric charge),
$q \bar q \gamma\gamma$ contact interactions of dimension-8. Searches
for such operators, with the
quarks replaced by electrons, have already been performed at TRISTAN and LEP
{\cite {aleph}} and have resulted in a lower bound of approximately 139 GeV
on the associated mass scale. As we will see below, the Tevatron(LHC) will
easily be able to push the scale in the corresponding $q \bar q$ situation
above 0.5-0.6(2.8-5.3) TeV.  As we will see below, unlike Higgs bosons or
other possible $\gamma \gamma$ resonances, this new contact interaction will
manifest itself only as a smooth modification of the various $\gamma \gamma$
distributions.

To be definitive, we will follow the notation employed by {\cite {con}} as
well as by the ALEPH Collaboration{\cite {aleph}} and assume that these
new interactions are parity conserving. (We will return to the cases where
such interactions may be chiral below.) In this case we can parameterize the
$q \bar q \gamma\gamma$ contact interaction as
\begin{equation}
{\cal L}= {2ie^2\over {\Lambda^4}}Q_q^2 F^{\mu\sigma}F_\sigma^\nu \bar q
\gamma_\mu\partial_\nu q  \,,
\end{equation}
where $e$ is the usual electromagnetic coupling, $Q_q$ is the quark charge, and
$\Lambda$ is the associated mass scale. The most obvious manifestation of this
new operator is to modify the conventional Born-level partonic $q \bar q \to
\gamma\gamma$ differential cross section so that it now takes the form
\begin{equation}
{d\hat \sigma\over {dz}}=Q_q^4 {2\pi \alpha^2\over {3\hat s}}
\left[{1+z^2\over {1-z^2}} \pm 2{\hat s^2\over {4\Lambda_{\pm}^4}}(1+z^2)+
\left({\hat s^2\over {4\Lambda_{\pm}^4}}\right)^2 (1-z^4)\right]  \,,
\end{equation}
where $\hat s,~z$ are the partonic center of mass energy and the cosine of
center of mass scattering angle, $\theta^*$, respectively. Note that we
have written $\Lambda_{\pm}$ in place of $\Lambda$ in the equation above
to indicate that the limits we obtain below will depend upon whether
the new operator constructively or destructively interferes with the SM
contribution. In our discussion, we focus on the case of constructive
interference  but will give results for both cases below.

There are two major
effects due to finite $\Lambda$: ($i$) Clearly, once $\hat s$
becomes comparable to $\Lambda^2$, the parton-level differential
cross section becomes less
peaked in the forward and backward directions implying that the photon pair
will generally be more central and will occur with higher average $p_t$'s.
($ii$) When integrated over
parton distributions the resulting cross section will lead to an increased
rate for photon pairs with large $\gamma\gamma$ invariant masses. From
these two observations we
see that the best hope for isolating finite $\Lambda$ contributions is to look
for excess diphotons with high, balanced $p_t$'s in the central detector
region with large pair masses.

Unfortunately, the $q \bar q \to \gamma\gamma$ tree-level
process is not the only one which produces diphotons satisfying the above
criteria. The issue of diphoton production at hadron colliders has been
discussed for many years{\cite {gamgam}}, mostly within the context of
searches for Higgs bosons in the intermediate mass region, and a full
next-to-leading(NLO) order SM calculation now exists{\cite {nlo}}. In the Higgs
search scenario, one is looking for a very narrow peak in the diphoton mass
distribution for which there are extremely large misidentification induced
backgrounds from QCD. In the case of contact interactions, however, the
deviation from SM expectations occurs over a rather wide range of diphoton
pair masses. The authors of
Ref.{\cite {nlo}} have provided an excellent summary of the various sources
which lead to diphoton pairs and we will generally follow their discussion.
The most obvious additional source of diphotons arises from the process
$gg\to \gamma\gamma$ which is induced by box diagrams. Although relatively
small in rate at the Tevatron, the increased $gg$ luminosity as one goes
to LHC energies (combined with the fact that $q\bar q$ annihilation is now a
`sea-times-valence' process) implies that $gg\to \gamma\gamma$ is extremely
important there. We include the $gg-$induced diphotons in our calculations by
employing 5 active quark flavors in the $gg$-induced box diagram for partonic
center of mass
energies, $\hat s$, below $4m_t^2$, 6 active flavors for $\hat s >>4m_t^2$,
and smoothly interpolate between these two cases. In order to include
the potential
effects of loop corrections to the rate for the $gg\to \gamma\gamma$ we have
scaled the results obtained by this procedure by an approximate `K-factor' of
1.3(1.5) at the Tevatron(LHC). (In numerical calculations, we assume
$m_t^{phys}=170$ GeV in agreement with precision electroweak measurements and
the recent evidence for top production at the Tevatron reported by the
CDF Collaboration{\cite {mor}}.) A similar `K-factor' is also employed in the
$q\bar q \to \gamma\gamma$ calculation; we use the results of
Ref.{\cite {vb}}. While this procedure gives only an approximate result to
the full NLO calculation, it is sufficient for our purposes as the effects of
the new contact interaction are quite large as they modify the tree level
cross section.

Additional `background' diphotons arise from three other sources.
For $2\to 2$ processes,
one can have either ($i$) single photon production through $gq \to \gamma q$
and $q\bar q\to g\gamma$ followed by the fragmentation $g,q \to \gamma$, or
($ii$) a conventional $2\to 2$ process with {\it both} final state $q,g$
partons fragmenting to photons. Although these processes appear to be
suppressed by powers of $\alpha_s$, these are off-set by large logs.
Numerically, in our analysis these fragmentation contributions are
calculated at the leading-log level using the results in
Refs.{\cite {nlo,frag}}. For $2\to 3$ processes, ($iii$) double
bremsstrahlung production of diphotons
is possible, \eg, $gq \to q\gamma \gamma$ or $q\bar q \to g\gamma \gamma$. All
of these `backgrounds' are relatively easy to drastically reduce or completely
eliminate at both the Tevatron and the LHC by the series of cuts we employ
below. We first
select diphotons with large invariant masses that are both central and
have high transverse momenta
($|\eta_{\gamma}|<1,~2.5$ and $p_t^{\gamma}>15,~100$ GeV at the Tevatron and
LHC respectively). We next demand that any additional hadronic activity in
these events be
widely separated (\ie, isolated) from the photons and have only small
associated jet energies
($E_j^{cut}<5,~20$ GeV at the Tevatron and LHC). As a last requirement
we demand that both photons are close to being back to back with nearly
balancing $p_t$'s, \ie, $p_t^1/(p_t^1+p_t^2)<0.7$, with $p_t^1 \geq p_t^2$ as
required by the ATLAS Higgs search analysis{\cite {ATLAS}}. It is easy to see
how these cuts will greatly reduce the backgrounds from the sources
($i$)-($iii$).  In process ($i$), the additional $\gamma$ produced from the
fragmentation essentially follows the fragmenting parton's direction but is
generally softer than the primary photon in the opposite hemisphere and
is contained within a jet which
has, on average, a relatively large energy. Under these circumstances, it is
very difficult for the two photons to be isolated and have their $p_t$'s
balance; it is also harder for the softer photon to pass the minimum $p_t$ cut
and
combine with the primary photon to pass the invariant pair mass requirements.
Process ($ii$) has an even more difficult time satisfying these cuts since the
photons
are both softer than their parent partons, are embedded in jets, and need to
have similar fragmentation energies to balance $p_t$'s. We find in practice
that the combination of the cuts above reduces the backgrounds from the sum
of sources ($i$) and ($ii$) by at least a factor $>15-20$ for both the Tevatron
and LHC for the diphoton pair masses of interest to us in our discussion
below.  In fact, for the machine luminosities we consider, these cuts
essentially eliminate ($ii$) as a diphoton background source.
For the $2\to 3$ processes ($iii$), most of the phase space is
dominated by the infrared and co-linear regions. Since {\it both} $\gamma$'s
must be hard and isolated with essentially matching $p_t$'s, the cuts above
are designed to remove this background source almost completely.

In presenting our numerical results, we follow the approach employed by the
CDF Collaboration in presenting their high $p_t$ diphoton data{\cite {cdf2}},
\ie, we integrate the invariant diphoton mass distribution above a given fixed
minimum value of the diphoton mass, $M_{\gamma\gamma}^{min}$, subsequent to
making all the other cuts discussed above. In order to get a feel for the
event rates involved, we scale this integrated cross section by a luminosity
appropriate to the Tevatron or the LHC, \ie, $20 pb^{-1}$ and $100 fb^{-1}$,
respectively. The results from this procedure are shown in Figs. 1 and 2.
Fig. 1a compares the SM diphoton cross section as a function of
$M_{\gamma\gamma}^{min}$ with the constructive interference scenario for
various values of the $\Lambda$ parameter. It is clear that present data
from the Tevatron{\cite {cdf2}} is already probing values of $\Lambda$ of
order 400 GeV. (In fact, CDF observes a possible excess of diphotons at
large values of $M_{\gamma\gamma}^{min}$ consistent with values of $\Lambda$
not far from 400 GeV. Of course, this is most likely a statistical
fluctuation.)

Assuming that no event excesses are observed, we can ask
for the limits that can be placed on $\Lambda_{\pm}$ as the Tevatron
integrated luminosity is increased. To do this we perform a Monte Carlo
study, first dividing the $M_{\gamma\gamma}^{min}$ range above 100 GeV into
nine steps of 50 GeV. Almost all of the sensitivity to finite $\Lambda$ will
lie in this range since for smaller values of $M_{\gamma\gamma}^{min}$ the
cross section looks too much like the SM while for larger values of
$M_{\gamma\gamma}^{min}$ the event rate is too small to be useful even for
integrated luminosities well in excess of $1 fb^{-1}$.  We generate events
using the SM as input and then try to fit the resulting
$M_{\gamma\gamma}^{min}$ distribution by a $\Lambda_{\pm}$-dependent fitting
function.
{}From this, bounds on $\Lambda$ are directly obtainable via a $\chi^2$
analysis. In the first approach, we assume that the normalization of the
cross section for small values of $M_{\gamma\gamma}^{min}$ using experimental
data will remove essentially all
of the systematic errors so that only statistical errors are put into the
fitting procedure. In this case, for a luminosity of $100(250,~500,~1000,
{}~2000) pb^{-1}$ we obtain the bounds $\Lambda_+>487(535,~575,~622,~671)$ GeV
and $\Lambda_->384(465,~520,~577,~635)$ GeV, respectively, at $95 \%$ CL.
Note that the constraints we obtain on $\Lambda_-$ are generally weaker than
those for $\Lambda_+$ although the two bounds begin to numerically converge at
larger integrated luminosities. If
we also allow for an {\it additional} overall systematic error of $15-20\%$
due to other uncertainties not accounted for by the normalization at small
$M_{\gamma\gamma}^{min}$ values, we find that our $\Lambda_+$ limits are
degraded by at most a factor of $\simeq 10\%$ while the corresponding ones for
$\Lambda_-$ are somewhat more significantly reduced, often by as much as
$20 \%$. Fig. 1b shows a sample case from this Monte Carlo
analysis wherein an integrated luminosity of 500 $pb^{-1}$ and only
statistical errors are assumed; the generated data together with the curves
associated with the two
$95 \%$ CL bounds and the best fit, corresponding to $\Lambda_-=1128$ GeV,
are displayed explicitly.

At the LHC, we find the results presented in Figs. 2a-c. Fig. 2a clearly
shows, for the default values of the cuts described above and an integrated
luminosity of $100 fb^{-1}$, that values of $\Lambda$ greater than 2 TeV will
be easily probed. If we strengthen our canonical $\gamma\gamma$ selection
cuts, \ie, $|\eta_{\gamma}|<1$ and $p_t^{\gamma}>200$ GeV, we see in Fig. 2b
that we
are left with reduced statistics but significantly improved sensitivity. If
we follow the same Monte Carlo approach as above, we obtain very strong
limits on $\Lambda_{\pm}$. We take the $M_{\gamma\gamma}^{min}$ range above
250 GeV and divide it into ten steps and generate data as above fitting to a
$\Lambda_{\pm}$-dependent distribution; we assume that the only errors are
statistical in this first pass as in the analysis above. (The best fit value
for the canonical cuts is found to be $\Lambda_+=5.17$ TeV.) From this
we obtain the $95\%$ CL
bounds of $\Lambda_+>2.83$ TeV and $\Lambda_->2.88$ TeV; the best fit and
the corresponding limit curves are shown in Fig. 2c. If we increase the
integrated
luminosity to $200 fb^{-1}$, the limits increase to $\Lambda_+>3.09$ TeV
and $\Lambda_->3.14$ TeV. If, instead, we strengthen the cuts, as
described above, we find these limits for a $100 fb^{-1}$ integrated
luminosity are improved to 5.27 TeV and 5.37 TeV,
respectively. If, in addition, we also simultaneously increasing the luminosity
we find instead the limits of 5.77 TeV
and 5.86 TeV, respectively. (These limits may be improved slightly by extending
the fit region outwards by another 500 GeV particularly in the higher
luminosity case.)

What happens when we give up the assumption of a parity conserving contact
interaction and assume that the quark current is chiral? The result is easily
obtained from the considerations of the ALEPH Collaboration{\cite {aleph}}.
Essentially, assuming either purely left- or right-handed quark currents, the
previously obtained values of $\Lambda_{\pm}$ are reduced by a factor of
$2^{1/4} \simeq 1.19$. This is not a sizeable effect numerically so that the
ranges of $\Lambda_{\pm}$ being probed are essentially unaltered.

In this paper, we have explored the capability of the Tevatron and LHC to
probe for the existence of flavor-independent contact interactions with the
following results:

($i$) If new dimension-8 $q\bar q \gamma\gamma$ contact interactions,
parameterized by a scale $\Lambda_{\pm}$, exist,
they will result in an excess of diphoton events with large invariant masses at
hadron colliders. These events will be central and the photons will have
large, balancing $p_t$'s. The resulting modifications to the $\gamma\gamma$
cross section for different $\Lambda$ at the Tevatron and LHC were obtained.
Existing constraints on the corresponding $e^+e^-$ operators are rather poor.

($ii$) While the new physics we propose only takes place in the
$q\bar q \to \gamma\gamma$
channel, there are other additional diphoton sources that must be accounted
for including the usual $gg \to \gamma\gamma$ box. We have estimated
their contributions and included them within the analysis. A large fraction
of these additional `backgrounds' are easily reduced by the cuts we have
imposed which simultaneously enriches the highly
central, large $p_t$, large $M_{\gamma\gamma}$ event sample which is most
sensitive to the new contact interaction.

($iii$) A Monte Carlo analysis was performed for both the Tevatron and LHC
to explore the potential constraints on $\Lambda_{\pm}$ for various integrated
luminosities assuming the absence of new physics. The generated data was fit
via a $\chi^2$ analysis to the $\Lambda_{\pm}$-dependent cross section thus
obtaining best fit values as well as $95\%$ CL limits under the assumption
that the cross section was properly normalized by the data at small values of
$M_{\gamma\gamma}$. For the Tevatron with ${\cal L}=500 pb^{-1}$, for
example, this led to limits in excess of 520 GeV and a best fit value of
$\Lambda_{-}=1128$ GeV. For the LHC with ${\cal L}=100 fb^{-1}$, the limits
were found to be sensitive to the choice of cuts and ranged from 2.8
to 5.3 TeV. Although the statistics were reduced by the stronger cuts, this
was more than compensated for by the increased sensitivity to the new contact
interaction.

Excess diphoton events should be searched for, not only as narrow peaks in
$M_{\gamma\gamma}$ signalling the existence of Higgs-like objects, but also
in the broad contributions to the tails of distributions. Such searches may
yield valuable information on the existence of new physics.

\vskip.25in
\centerline{ACKNOWLEDGEMENTS}

The author would like to thank R. Blair of the CDF Collaboration as well as
J.\ Ohnemus and J.L.\ Hewett for discussions related to this work. The author
would also like to thank the members of the Argonne National Laboratory High
Energy Theory Group for use of their computing facilities and the High Energy
Physics Group at the University of Hawaii, where part of this work was done,
for their hospitality.

\newpage

%
%%%%%%%%%%%%%%%%%%--- References
%%%%%%%%%%%%%%%%%%%%%%%%%%%%%%%%%%%%%%%%%%%%%%%%%%%%%%%
\def\MPL #1 #2 #3 {Mod.~Phys.~Lett.~{\bf#1},\ #2 (#3)}
\def\NPB #1 #2 #3 {Nucl.~Phys.~{\bf#1},\ #2 (#3)}
\def\PLB #1 #2 #3 {Phys.~Lett.~{\bf#1},\ #2 (#3)}
\def\PR #1 #2 #3 {Phys.~Rep.~{\bf#1},\ #2 (#3)}
\def\PRD #1 #2 #3 {Phys.~Rev.~{\bf#1},\ #2 (#3)}
\def\PRL #1 #2 #3 {Phys.~Rev.~Lett.~{\bf#1},\ #2 (#3)}
\def\RMP #1 #2 #3 {Rev.~Mod.~Phys.~{\bf#1},\ #2 (#3)}
\def\ZP #1 #2 #3 {Z.~Phys.~{\bf#1},\ #2 (#3)}
\def\IJMP #1 #2 #3 {Int.~J.~Mod.~Phys.~{\bf#1},\ #2 (#3)}

\newpage

%%%%%%%%%%%%%%%%%%%%%%%--- figures
%
{\bf Figure Captions}
\begin{itemize}

\item[Figure 1.]{(a)Diphoton pair event rate, scaled to an integrated
luminosity of $20 pb^{-1}$, as a function of $M_{\gamma\gamma}^{min}$ at the
Tevatron subject to the cuts discussed in the text. The solid curve is the
QCD prediction, while from top to bottom the dash dotted curves correspond to
constructive interference with the SM and
a compositeness scale associated with the $q\bar q \gamma\gamma$ operator of
$\Lambda_+=0.2,~0.3,~0.4,~0.5,$ and 0.6 TeV respectively. (b) Monte Carlo
generated data for an integrated luminosity of 500 $pb^{-1}$ assuming
statistical errors only. The two dash dotted curves corresponding to the
$95 \%$ CL bounds discussed in the text while the solid curve represents the
best fit.}
\item[Figure 2.]{Same as Fig. 1, but for the LHC scaling to an integrated
lumonosity of $100 fb^{-1}$. From top to bottom the
dash dotted curves now correspond to $\Lambda_+=0.75,~1.0,~1.25,~1.5,~1.75$ and
2.0 TeV respectively. In (a) we require $p_t^{\gamma}>100$ GeV and
$|\eta_\gamma|<2.5$, while in (b) we require instead $p_t^{\gamma}>200$ GeV
and $|\eta_\gamma|<1.0$. (c) Monte Carlo
generated data for an integrated luminosity of 100 $fb^{-1}$ assuming
statistical errors only. The two dash dotted curves corresponding to the
$95 \%$ CL bounds discussed in the text while the solid curve represents the
best fit.}
\end{itemize}

\end{document}